\documentclass[conference]{IEEEtran}
\IEEEoverridecommandlockouts
\usepackage{cite}
\usepackage{amsmath,amssymb,amsfonts}
\usepackage{algorithmic}
\usepackage{graphicx}
\usepackage{textcomp}
\usepackage{xcolor}
\def\BibTeX{{\rm B\kern-.05em{\sc i\kern-.025em b}\kern-.08em
    T\kern-.1667em\lower.7ex\hbox{E}\kern-.125emX}}
\begin{document}

\title{Parameterization of Sequence of MFCCs for DNN-based voice disorder detection\\
}

\author{\IEEEauthorblockN{Tomasz Grzywalski}
\IEEEauthorblockA{\textit{StethoMe\textsuperscript{\textregistered}} \\
Poznan, Poland \\
grzywalski@stethome.com}
\and
\IEEEauthorblockN{Adam Maciaszek}
\IEEEauthorblockA{\textit{StethoMe\textsuperscript{\textregistered}} \\
Poznan, Poland \\
maciaszek@stethome.com}
\and
\IEEEauthorblockN{Adam Biniakowski}
\IEEEauthorblockA{\textit{StethoMe\textsuperscript{\textregistered}} \\
Poznan, Poland \\
biniakowski@stethome.com}
\and
\IEEEauthorblockN{Jan Orwat}
\IEEEauthorblockA{\textit{StethoMe\textsuperscript{\textregistered}} \\
Poznan, Poland \\
orwat@stethome.com}
\and
\IEEEauthorblockN{Szymon Drgas}
\IEEEauthorblockA{\textit{StethoMe\textsuperscript{\textregistered}} \\
Poznan, Poland \\
drgas@stethome.com}
\and
\IEEEauthorblockN{Mateusz Piecuch}
\IEEEauthorblockA{\textit{StethoMe\textsuperscript{\textregistered}} \\
Poznan, Poland \\
piecuch@stethome.com}
\and
\IEEEauthorblockN{Riccardo Belluzzo}
\IEEEauthorblockA{\textit{StethoMe\textsuperscript{\textregistered}} \\
Poznan, Poland \\
belluzzo@stethome.com}
\and
\IEEEauthorblockN{Krzysztof Joachimiak}
\IEEEauthorblockA{\textit{StethoMe\textsuperscript{\textregistered}} \\
Poznan, Poland \\
joachimiak@stethome.com}
\and
\IEEEauthorblockN{Dawid Niemiec}
\IEEEauthorblockA{\textit{StethoMe\textsuperscript{\textregistered}} \\
Poznan, Poland \\
niemiec@stethome.com}
\and
\IEEEauthorblockN{Jakub Ptaszynski}
\IEEEauthorblockA{\textit{StethoMe\textsuperscript{\textregistered}} \\
Poznan, Poland \\
ptaszynski@stethome.com}
\and
\IEEEauthorblockN{Krzysztof Szarzynski}
\IEEEauthorblockA{\textit{StethoMe\textsuperscript{\textregistered}} \\
Poznan, Poland \\
szarzynski@stethome.com}
}

\maketitle

\begin{abstract}
In this article a DNN-based system for detection of three common voice disorders (vocal nodules, polyps and cysts; laryngeal neoplasm; unilateral vocal paralysis) is presented. The input to the algorithm is (at least 3-second long) audio recording of sustained vowel sound /a:/. The algorithm was developed as part of the "2018 FEMH Voice Data Challenge" organized by Far Eastern Memorial Hospital and obtained score value (defined in the challenge specification) of 77.44. This was the second best result before final submission. Final challenge results are not yet known during writing of this document. The document also reports changes that were made for the final submission which improved the score value in cross-validation by 0.6\% points.
\end{abstract}


\section{Introduction}
{\let\thefootnote\relax\footnotetext{\textcopyright 2018 IEEE}}

Common voice pathologies include structural lesions, neoplasms, and neurogenic disorders. These disorders are typically diagnosed using laryngeal endoscopy. 
Non-invasive screening methods, based on analysis of acoustic recordings, provide means to assess the health of vocal apparatus without expensive equipment operated by well-trained specialists. 

The detection of voice pathologies can be performed from recordings of sustained vowels or continuous speech \cite{parsa2001acoustic}. While continuous speech contains variations of pitch and loudness which can be important for detection of voice pathologies, sustained vowels provide more reliable parameters, as characteristics of the voice source, vocal tract, and articulators are relatively time-invariant. In contrast to continuous speech, for sustained vowels individual speech characteristics such as speaking rate, speaker's dialect does not affect the detection. The use of sustained vowel samples is expected to generate a simpler acoustic structure that might lead to consistent and reliable decisions.

In many works in the literature about automatic voice pathology detection from sustained vowels, acoustic features are extracted which are next classified using various machine learning methods. Most of the acoustic features are related to perturbations like jitter and schimmer \cite{teixeira2014jitter}. As it may be difficult to extract fundamental frequency for pathological voice, a feature which is needed to extract jitter, it was proposed in \cite{arias2011combining,markaki2011voice} to extract features based on modulation spectrum. 

\begin{figure*}[!th]
\centering
\includegraphics[width=16cm]{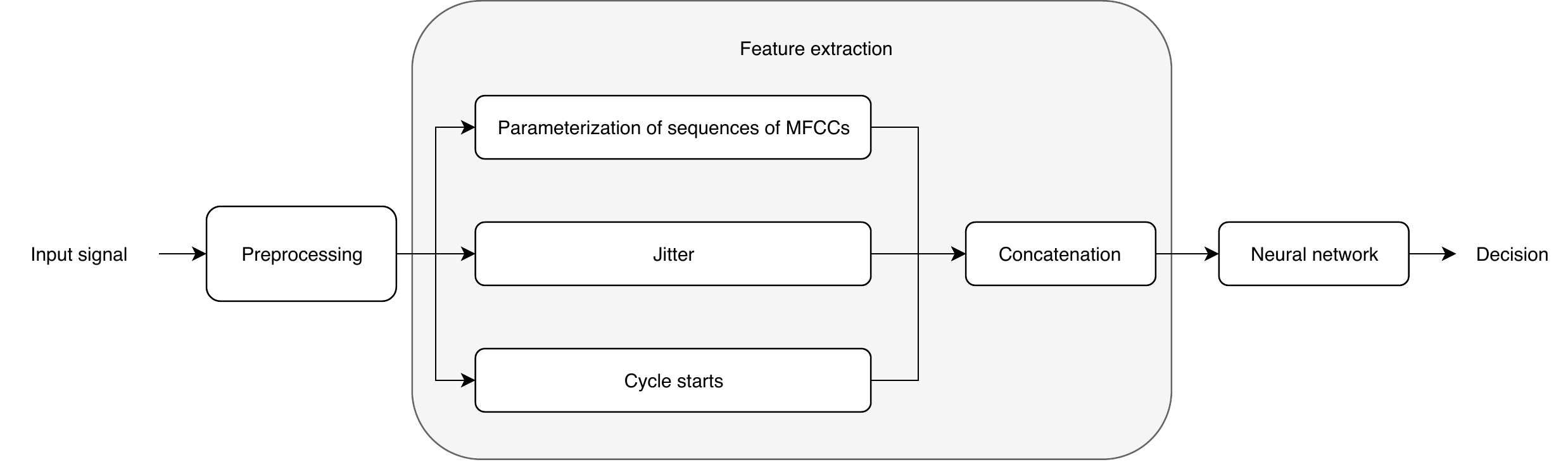}
\caption{Overview of proposed solution}
\label{overall_system}
\end{figure*}

Another class of features broadly used in voice pathology detection and classification contains compact representations of short-time spectra of speech like MFCCs (mel-frequency cepstral coefficients) or LPC (linear prediction coding). This kind of features was used in \cite{FANG2018}. Distributions of such features can be modeled using GMMs (Gaussian mixture models), or using neural networks where each feature vector is classified individually, and classification scores are combined, for example by averaging.

In \cite{strathprints64290} convolutive neural network was trained to classify whole spectrograms. The Experiments were performed on Saarbruecken voice database, which contains recordings from 2000 individuals. 

Many voice pathology detection systems known from the literature, detection is based on 
statistics of perturbations parameterized locally. However, some temporal information is used in detection when HMMs (hidden Markov models) are applied \cite{mesallam2017development}.

In this work, we propose features, which carry information about time trends of cepstral coefficients representing spectra in a compact way, i.e. time sequences of MFCCs are modeled by polynomials. The polynomial coefficients are concatenated with jitter and cyclestarts features \cite{hadjitodorov2002computer}. This approach can carry information about onset and offset of the sustained vowel. Such a compact representation allowed us to apply deep neural network for FEMH database.

\section{Proposed method}
\subsection{General architecture}

The overall architecture of our proposed system is depicted in Figure \ref{overall_system}, it consists of the following steps:
\begin{enumerate}
    \item \emph{Preprocessing} - The purpose of this pre-processing stage is to extract each separate vowel sound from recordings. Few recordings in the dataset included couple attempts made by the patient to pronounce the vowel. We want to extract each separate sound and precisely crop it so that beginnings of vowel sounds align. All further operations are performed individually for each vowel sound. If a recording contained multiple sounds, in the test phase each sound was individually classified and class scores were averaged.
    \item \emph{Feature extraction} - In this stage from each vowel sound a feature vector is extracted. 
    In the described system three types of features are extracted and the resulting vectors are concatenated into one feature vector. In this paper, we propose to model sequences of MFCCs as described in sections \ref{sec:tcmod1} and \ref{sec:tcmod2}. Feature vector resulting from the proposed extractor is concatenated with features widely used for detection of pathologies: cyclestarts \cite{hadjitodorov2002computer} and jitter \cite{teixeira2014jitter}.

    \item \emph{Neural network} - Our classifier of choice is neural network that accepts on input a set of features and produces a probability of the sound belonging to one of the four defined classes.
\end{enumerate}

\begin{figure}
    \centering
    \includegraphics[width=9cm]{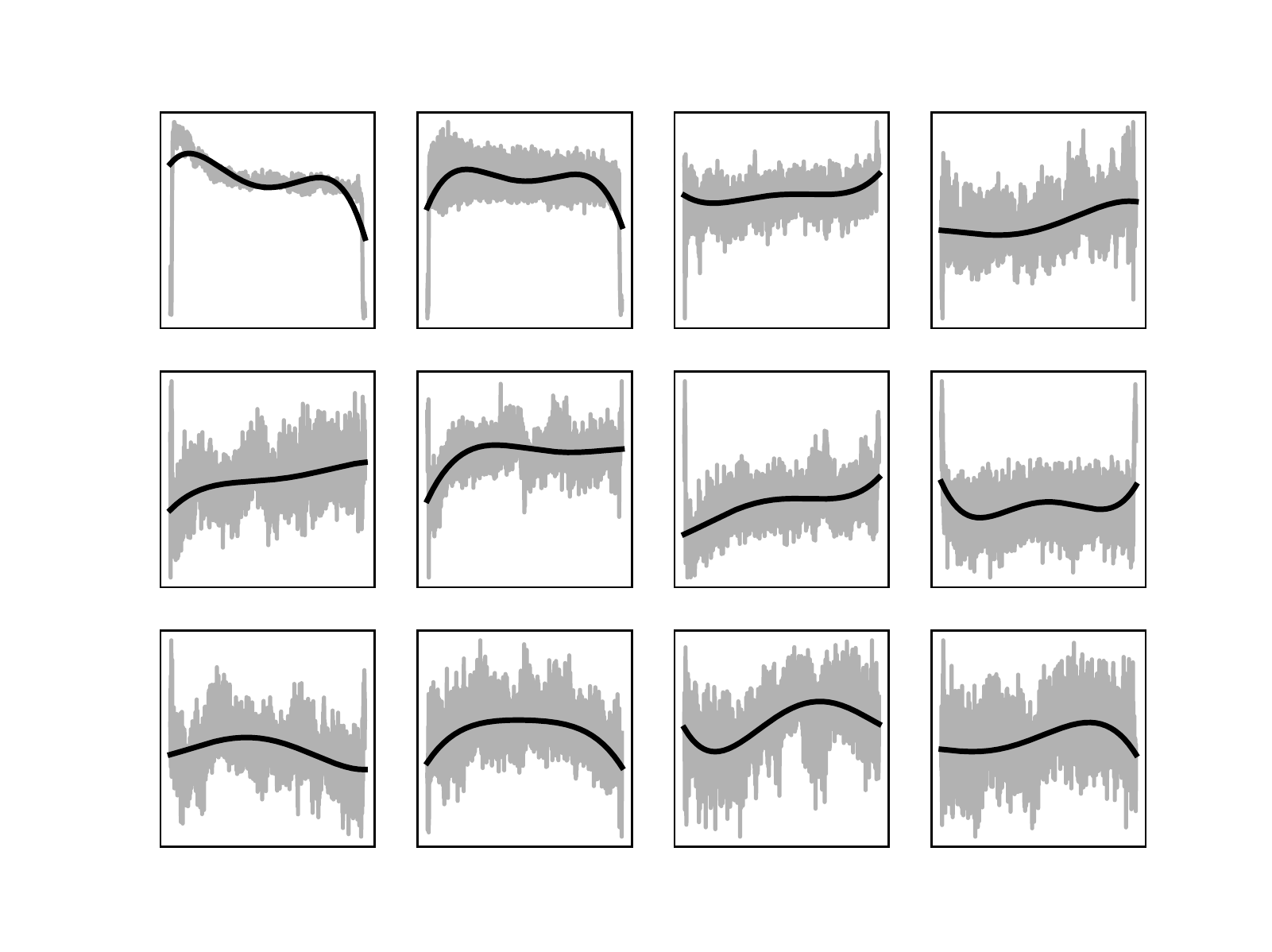}
    \caption{Examples that show polynomials which approximate how individual MFCCs change over time.}
    \label{fig:pol}
\end{figure}

\subsection{Modeling sequences of cepstral coefficients with splines}
\label{sec:tcmod1}
In order to provide a compact representation of the evolution of short-time spectra during realization of sustained vowel, we propose to model sequences of MFCCs with polynomials. Let denote $i$'th MFCC in $j$'th frame as $c_{ij}$, where $i=1,\ldots,I$ and $j=1,\ldots,J$. For each of $I$ sequences of points 
\begin{equation}
\label{eq:seq}
S_i=((1, c_{i1}),\ldots, (N, c_{iN}))\;,  
\end{equation}
 where $N$ is the number of points used in the analysis, a polynomial is fitted using least squared error criterion. Two variants of polynomial fitting are proposed. In the first one all points were used ($N=J$).

 We have observed that patients with vocal pathologies tend to have much longer onset of a vowel, therefore we designed a features that represent only the beginning of the vowel sound. In this case, $N<J$ and MFCCs from about 50 first frames are used. 
 In this case spline with explicit $K$ internal knots is fitted. The first coordinates of internal knots $k_1,\ldots,k_K$ are distributed uniformly from $0$ to $N$. Finally, values of the fitted spline for arguments $k_1,\ldots,k_K$ form a feature vector. 
 
\subsection{MFCC-FFT}
\label{sec:tcmod2}
Previous two set of features described the whole sample length and very beginning of the sample. To complete this description we introduce a third set of features that captures characteristics from the initial fragment of the vowel example but with the length equal to the average length of vowel examples in our dataset (that is about two seconds). In this case, instead of using polynomials to model time courses of MFCCs, FFT was computed from each sequence  $(c_{i1},\ldots,c_{iN})$ for $i=1,\ldots,I$, and absolute values of its first coefficients were used as features. Thus, in contrast to the polynomial-based features, only rates of changes of MFCCs are represented but not the shapes of time courses.

\section{Evaluation}
\subsection{Data}
Data provided for the challenge consisted of training dataset comprising 200 recordings with known labels of 4 classes (normal; vocal nodules, polyps and cysts; laryngeal neoplasm and unilateral vocal paralysis) and test set comprising 400 recordings with unknown labels. Predicting correct labels for the 400 test recordings was the aim of the challenge.

Number of recording for each of the four classes is presented in Table \ref{train_stat}.

\begin{table}[htbp]
\caption{Training set statistics}
\begin{center}
\begin{tabular}{|c|c|c|c|c|}
\hline
Class                            & Normal & Neoplasm & Phonotrauma & Vocal palsy \\
\hline
Num. of recordings               & & & &\\
in train set                     & 50     & 40       & 60          & 50 \\
\hline
\end{tabular}
\label{train_stat}
\end{center}
\end{table}

Each recording was a single-channel audio file with 44100 Hz sampling rate.

\subsection{Cropping}
\label{sec:crop}
The first stage of data preprocessing was extraction of separate vowel sounds. We determined when the signal exceeds a particular volume level and extracted time excerpts in which the sound is present, including an offset of 0.15s before and 0.25s after the determined signal excerpt. Detailed algorithm for cropping is following:

\begin{enumerate}
    \item Low-pass filtering of the input signal at 1000 Hz using 6th order IIR Butterworth filter
    \item Computation of Hilbert envelope from signal
    \item Filtering the envelope (moving average, window length of 1100 samples)
    \item Computation of signal level $L$ over time from the filtered envelope $E(t)$ according to equation: $L(t) =20log10(E(t))$ where $t$ is the index of a sample.
    \item Compute threshold level $Lth =(L2+L95)/2 + 3 dB$ where $L2$ and $L95$ are 2 and 95 percentiles of distribution of signal level $L$
    \item Get fragments of signals where level is above $Lth$ and remove fragments shorter than 500 ms
    \item Merge fragments if they are separated by less than 200ms 
    \item Cut the input signal according to the detected fragments extended 150 ms and 250 ms before and after each fragment
\end{enumerate}

After this stage we obtained 213 vowel sounds from 200 training recordings.

\subsection{Data augmentation}
For every vowel sound extracted as described in Section \ref{sec:crop}, we created three data augmentations which were used alongside the original example during model training. Each augmentation was an original vowel sound with randomly applied: volume level change, pitch shift and time stretch. The parameters of the modifiers were following:
\begin{itemize}
    \item volume change by simple multiplication by a random value uniformly sampled from range 0.4 to 1.2,
    \item random pitch shift drawn from normal distribution with mean equal 0.0 and standard deviation equal 0.5,
    \item time stretch with stretch rate sampled uniformly from range 0.85 to 1.5.
\end{itemize}
After this stage we gathered 852 training samples including original and augmented ones.

\subsection{Feature extraction}
\subsubsection{MFCC}
Part of elements of feature vector were computed from mel-frequency cepstrum (MFCC) coefficients. Parameters of the MFCC extraction were following:
\begin{itemize}
    \item \emph{Preemphasis filter with coefficient = 0.97}
    \item \emph{Window length = 0.008s}
    \item \emph{Window step = 0.011s}
    \item \emph{FFT size (n\_fft) = 512}
    \item \emph{Number of mel filters = 26}
    \item \emph{Number of returned MFCCs = 13}
\end{itemize}
Provided parameters were selected after a series of experiments where window lengths from 0.0027s to 0.019s and windows steps from 0.0027s to 0.044s were tested in a grid-search manner. Based on MFCC extracted using these parameters, two main sets of features were extracted from each sample:

\begin{itemize}
    \item \emph{mfcc\_polynomials}: For  each sequence of the 13 MFCCs a polynomial of 4-th degree was fitted as described in Section \ref{sec:tcmod1}. This resulted with $13\times5=65$ features. Examples of obtained polynomials are depicted in Figure \ref{fig:pol}
    \item \emph{mfcc\_splines}: Splines were fitted to the sequences of MFCCs from the first 50 frames (about 544ms of recording length) (see Section \ref{sec:tcmod2}). In case the sample was shorter - last frame was repeated till a total length of 50 was obtained. Knots of the splines were fixed at 0th, 10th, 20th, 30th, 40ht, 50th element of each modeled sequence. Next, approximated values of modeled sequence in splines knots locations were extracted. There were a total of $13\times6=78$ such features extracted this way from each sample.

    \item \emph{mfcc\_fft}:  Similar as in mfcc\_splines we parameterize MFCCs from the first 150 frames. In this case, however, parameterization is done by means of fast Fourier transform (FFT). From each obtained magnitude spectrum of size 76 (half of spectrum which is symmetric), the first 6 coefficients are extracted which becomes a set of $13\times6=78$ features. 
\end{itemize}

\subsubsection{Jitter and cycle starts}
Additional to features extracted from MFCC we also included two other set of features that we hoped would capture different characteristics of the vowel sounds:
\begin{itemize}
    \item \emph{jitter}: Five types of jitter parameters have been extracted as in \cite{teixeira2014jitter}: absolute jitter, relative average perturbation ($rap$), the 5-point period pertubation quotient ($ppq5$), and the difference of differences of periods ($ddp$). Additionally, we extract F0 and HNR. Finally a feature vector from this extractor has 7 elements.
    \item \emph{cyclestarts}: Comprises of two features: Turbulent Noise Index (TNI) and Normalized First Harmonic Energy (NFHE) as defined in \cite{hadjitodorov2002computer} (2 values).
\end{itemize}

\subsection{Neural network}
Our classifier was a neural network containing 2 hidden layers and an output layer. All hidden layers and the output layer were fully-connected. Each hidden layer featured 128 neurons with Exponential Linear Units (ELUs) nonlinearities \cite{clevert2015fast} and batch normalization \cite{ioffe2015batch}. Output layer contained 4 neurons with softmax nonlinearity so that the output values could be interpreted as probability.

To reduce over-fitting in training phase we used dropout \cite{srivastava2014dropout} before every layer in our network. The dropout rates before first, second and third layer were accordingly: 55\%, 25\% and 10\%.

All mentioned parameters were subject to manual optimization. Among tested options were: different number of hidden layers (2 and 3), different number of neurons in hidden layers (from 32 to 512), hidden layer's nonlinearities and presence/absence of batch normalization.

\subsection{Training}
\label{sec:train}
We split the dataset using 5-fold crossvalidation into three parts: training set (64\%), validation set (16\%), test set (20\%). We repeated this splitting process 20 times using different random seeds (we call each repetition an experiment). Training sets were used for networks training. By measuring trained network accuracy on validation set after each epoch, we selected the best model version for given seed and fold. Test sets were used for final performance evaluation, the results were aggregated across all folds and averaged across all 20 experiments. For the submission we used all 5 models from 20 experiments (100 models in total) and averaged obtained class probability scores.

All models were trained to minimize categorical crossentropy between network's predictions and target labels. Optimization was performed using Adam algorithm \cite{kingma2014adam}. Each fold was trained for 100 epochs with batch size of 32. Initial learning rate was 0.001 and was multiplied by weight decay factor of 0.985 after each epoch. Initially in one epoch each vowel sound from train set was used once, but starting from mid-October submission we used "Normal" sounds two times in each epoch to account for the fact that the train set contained fewer normal sounds (50) compared to pathological (150) \cite{chawla2009data}.

\subsection{Scoring}
All challenge submissions were evaluated and ranked using "score" value defined as weighted average of Sensitivity, Specificity and Unweighted Average Recall (UAR) with respective weights: 0.4, 0.2 and 0.4. Sensitivity and Specificity were defined for 2-class classification (normal vs any pathology) while UAR described model's ability to distinguish between three pathologies (neoplasm, phonotrauma and vocal palsy).

\subsection{Evaluated models}
During the competition we have submitted number of models for evaluation. Three of them, including our best performing model and final submission, are different variations of the system described in previous sections. The details of the three models are following:
\begin{enumerate}
    \item \emph{August submission}: This version used mfcc\_polynomials, jitter and cycle starts (65 + 7 + 2 = 74) features. Dropout before first layer was set to 10\% and "Normal" sounds appeared only once in each epoch
    \item \emph{mid-October submission}: This version used mfcc\_polynomials, mfcc\_splines, jitter and cycle starts (65 + 78 + 7 + 2 = 152) features and dropout of 55\% before first layer
    \item \emph{final submission}: This version used mfcc\_polynomials, mfcc\_splines and mfcc\_fft (65 + 78 + 78 = 221) and dropout of 60\% before first layer
\end{enumerate}

\subsection{Results}
Table \ref{tab1} contains results obtained by different versions of our system using 5-fold cross-validation procedure described in section \ref{sec:train}. Table \ref{tab2} contains results on the test set as reported by the Challenge organizers.

\begin{table}[htbp]
\caption{Cross-validation results}
\begin{center}
\begin{tabular}{|c|c|c|c|c|}
\hline
Submission        & Sensitivity & Specificity & UAR     & Score \\
\hline
August            & 89.9\%      & 80.5\%      & 67.47\% & 79.03 \\
\hline
mid-October       & 90.7\%      & 86.9\%      & 60.63\% & 77.91 \\
\hline
final             & 92.0\%      & 85.9\%      & 62.00\% & 78.52 \\
\hline
\end{tabular}
\label{tab1}
\end{center}
\end{table}

\begin{table}[htbp]
\caption{Test set results}
\begin{center}
\begin{tabular}{|c|c|c|c|c|}
\hline
Submission        & Sensitivity & Specificity & UAR     & Score \\
\hline
August            & 93.2\%      & 50.0\%      & 72.63\% & 76.33 \\
\hline
mid-October       & 89.4\%      & 66.0\%      & 71.20\% & 77.44 \\
\hline
\end{tabular}
\label{tab2}
\end{center}
\end{table}

\section{Conclusions}
In this article a system for detection of voice disorders was presented. Developed as part of the FEMH Voice data Challenge 2018, the system achieved second best score value (77.44) across all pre-final model evaluation rounds. In our final submission we have introduced changes that improved our results in cross-validation by 0.6 percent points.



\bibliographystyle{plain}
\bibliography{refs}

\end{document}